\documentclass[fleqn,usenatbib]{mnras}

\usepackage{newtxtext,newtxmath}
\usepackage{mathrsfs}
\usepackage{multirow}

\usepackage[T1]{fontenc}
\usepackage{ae,aecompl}

\usepackage{graphicx}	
\usepackage{amsmath}	

\usepackage{amssymb}	

\title[A Dyson Sphere around a black hole]{A Dyson Sphere around a black hole}

\author[Hsiao et al. 2020]{Tiger Yu-Yang Hsiao,$^{1}$
Tomotsugu Goto,$^{1}$
Tetsuya Hashimoto,$^{1,2}$
Daryl Joe D. Santos,$^{1}$
\newauthor
Alvina Y. L. On,$^{1,2,3}$
Ece Kilerci-Eser,$^{4}$
Yi Hang Valerie Wong,$^{1}$
Seong Jin Kim,$^{1}$
\newauthor
Cossas K.-W. Wu,$^{5}$
Simon C.-C. Ho,$^{1}$
Ting-Yi Lu,$^{1}$
\\
$^{1}$Institute of Astronomy, National Tsing Hua University, 101, Section 2. Kuang-Fu Road, Hsinchu, 30013, Taiwan (R.O.C.)\\
$^{2}$Centre for Informatics and Computation in Astronomy (CICA), National Tsing Hua University,101, Section 2. Kuang-Fu Road, Hsinchu, 30013, Taiwan (R.O.C.)\\
$^{3}$Mullard Space Science Laboratory, University College London, Holmbury St Mary, Surrey RH5 6NT, UK\\
$^{4}$Sabanc{\i} University, Faculty of Engineering and Natural Sciences, 34956, Istanbul, Turkey\\
$^{5}$Department of Physics, National Tsing Hua University, 101, Section 2. Kuang-Fu Road, Hsinchu, 30013, Taiwan (R.O.C.)
}

\date{Accepted 2021 June 25. Received 2021 June 25; in original form 2021 February 08}

\pubyear{2021}

\begin{document}
\label{firstpage}
\pagerange{\pageref{firstpage}--\pageref{lastpage}}
\maketitle
\begin{abstract}

The search for extraterrestrial intelligence (SETI) has been conducted for nearly 60 years.
A Dyson Sphere, a spherical structure that surrounds a star and transports its radiative energy outward as an energy source for an advanced civilisation, is one of the main targets of SETI.
In this study, we discuss whether building a Dyson Sphere around a black hole is effective.
We consider {six} energy sources: (i) the cosmic microwave background, (ii) the Hawking radiation, (iii) an accretion disk, {(iv) Bondi accretion, (v) a corona, and (vi) relativistic jets.}
To develop future civilisations (for example, a Type II civilisation), {$4\times10^{26}\,{\rm W}$($1\,{\rm L_{\odot}}$)} is expected to be needed.
Among (iii) to {(vi)}, the largest luminosity can be collected from an accretion disk, reaching 
{$10^{5}\,{\rm L_{\odot}}$,} enough to maintain a Type II civilisation.
Moreover, if a Dyson Sphere collects not only the electromagnetic radiation but also other types of energy (e.g., kinetic energy) from the jets, the total collected energy would be approximately 5 times larger.
Considering the emission from a Dyson Sphere, our results show that the Dyson Sphere around a stellar-mass black hole in the Milky Way ($10\,\rm kpc$ away from us) is detectable in the ultraviolet$(\rm 10-400\,{\rm nm)}$, optical$(\rm 400-760\,{\rm nm)}$, near-infrared($\rm 760\,{\rm nm}-5\,{\rm \mu m}$), and mid-infrared($\rm 5-40\,{\rm \mu m}$) wavelengths via the waste heat radiation using current telescopes such as {Galaxy Evolution Explorer Ultraviolet Sky Surveys.
Performing model fitting to observed spectral energy distributions and measuring the variability of radial velocity may help us to identify these possible artificial structures.}

\end{abstract}


\maketitle


\section{Introduction}
\label{introduction}

Since the very early stage of human history, our ancestors have already imagined the existence of aliens.
Numerous science fiction movies, television series, and fiction have demonstrated extraterrestrial intelligence in different forms.
However, even to this day, the debate on the existence of aliens or extraterrestrial intelligence in reality continues.
Despite the controversy, the search for extraterrestrial intelligence (SETI) has been conducted since the 1960s (Project Ozma)\footnote{\url{https://www.seti.org/seti-institute/project/details/early-seti-project-ozma-arecibo-message}}. 
Fortunately, after around 30 years since SETI began, the first exoplanet (Gamma Cephei Ab) was discovered \citep{Campbell1988}, and a new window was opened for the search for extraterrestrial life.
In the next decade, the James Webb Space Telescope (JWST) plans to study the atmosphere of the exoplanet.
Hence, JWST has a potential to reveal the origin of life and find extraterrestrial intelligence\footnote{\url{https://www.jwst.nasa.gov/content/science/origins.html}}.
Although there has hitherto been negative results, it never diminishes human curiosity in SETI. 

Many studies discussed and provided different concepts related to the advanced civilisations \citep[e.g.,][]{Dyson1960,Shkadov1988,Lee2013}.
One of the most famous concepts is the "Dyson Sphere", which was first described by \citet{Dyson1960}.
\citet{Dyson1960} argued that if there was an advanced civilisation in the Universe, it must consume extremely high energy to maintain itself.
In such a civilisation, its parent star (the star which its planets orbit around) could be a candidate to provide ample energy for development and sustainability.
Assuming that the highly-developed civilisation is eager to absorb the energy from its sun, a Dyson Sphere is a spherical structure that fully surrounds the star and directs all the radiative energy outward.
For instance, if a Dyson sphere is built around the Sun, the total luminosity of the Sun ($\rm{L_{\odot}\sim4\times10^{26}\,{\rm W}}$) can be utilised, which is approximately nine orders of magnitude larger than the power intercepted by the Earth ($\sim1.7\times10^{17}\,{\rm W}$).
After receiving the energy from the star, the civilisation would be able to convert the energy from low-entropy to high-entropy and emanate the waste heat (e.g., in mid-infrared wavelengths) into the background, suggesting this kind of energy waste is detectable \citep{Dyson1960}.

However, building a rigid Dyson Sphere is almost mechanically impossible due to the gravity and the pressure from the central star \citep[e.g.,][]{Wright2020}.
Hence, some variants of the concept were proposed.
For instance, a Dyson Swarm\footnote[3]{\label{fn3}\url{https://www.aleph.se/Nada/dysonFAQ.html}} is a group of collectors that orbits the central energy source \citep[see also][]{Dyson1960b}.
This method allows the civilisation to grow incrementally.
Nevertheless, due to the orbital mechanics, the arrangement of such collectors would be extraordinarily complex.
Another variant is a Dyson Bubble$^{\ref{fn3}}$ which contains a host of collectors as well.
However, the collectors are instead stationary in space assuming equilibrium between the outward radiative pressure and the inward gravity, which are usually called "solar sails" or "light sails".

Based on energy consumption, \citet{Kardashev1964} classified the hypothetical civilisations into three categories (Kardashev scale).
A Type I civilisation uses infant technology and consumes $4\times10^{16}\,{\rm W}$ (or $4\times10^{23}\,{\rm erg\ s^{-1}}$)\footnote[4]{In \citet{Kardashev1964}'s original definition, a Type I civilisation consumes an energy of that close to a planet, i.e., $4\times10^{12}\,{\rm W}$. Nowadays, a Type I civilisation is usually defined as harvesting all the power of a planet.}, the energy of its planet.
A Type II civilisation harvests all the energy of its parent star, namely $4\times10^{26}
\,{\rm W}$ (or $4\times10^{33}\,{\rm erg\ s^{-1}}$).
A Type III civilisation represents the highest technological level, which can engulf its entire galaxy as its energy source.
A typical energy used by a Type III civilisation is about $4\times10^{37} \,{\rm W}$ (or $4\times10^{44}\,{\rm erg\ s^{-1}}$).
To represent the civilisations that have not been able to use all of their available energy sources,  \citet{Sagan1973} suggested a logarithmic interpolation in the form of $K=0.1(\rm{log}\it{P}-6)$, where $K$ is the Kardashev index and $P$ (in the unit of Watts) indicates the energy consumption.
Currently, our civilisation level is approximately 0.73\footnote[5]{According to the world energy report of 2020: \url{https://www.bp.com/content/dam/bp/business-sites/en/global/corporate/pdfs/energy-economics/statistical-review/bp-stats-review-2020-full-report.pdf}, the total world energy consumption in 2019 is 583.9 exajoule ($\sim1.85\times10^{13}W$).}.
We may become a Type I civilisation in 200 to 800 years, a Type II in 1000 to 3000 years, and a Type III in 2000 to 5000 years, assuming the energy consumption increases by $1\%-3\%$ per year \citep[e.g.,][]{Gray2020}. 

The purpose of this study is to discuss the possibility of black holes being a candidate for an energy source to build a Dyson Sphere of a civilisation between a Type II and a Type III ($2<K<3$; a Type II civilisation).
As we mentioned in the previous paragraph, after a Type II civilisation absorbs all the energy from the parent star, a Type II civilisation would seek another star to maintain itself.
We would like to answer the question if black holes can be regarded as proper energy sources, or if they are inefficient to provide ample energy for civilisations to thrive.
\citet{Inoue2011} discussed a similar idea by describing a Dyson Sphere around a supermassive black hole (SMBH).
They assumed the energy from the accretion disk of a SMBH as the only energy source for a Type III civilisation. Their Dyson Sphere (collectors of power plants) is located between the accretion disk and the molecular torus, and is built to partly cover a SMBH to avoid absorbing the piercing jets to melt the Dyson Sphere.
However, the possibility to detect these structures is extremely low \citep[the chance to detect an unexpected signal from $1\,{\rm {\mu}arcsecond}$ is less than $10^{-23}$;][]{Inoue2011}.

Moreover, \citet{Opatrny2017} discussed a Dyson Sphere which absorbs the energy from the cosmic microwave background (CMB).
Compared with the CMB, a black hole should be a "cold" sun, turning the idea of the Dyson Sphere inside out (Here we name it as the Inverse Dyson Sphere; IDS). 
Based on their results, a black hole with one solar mass could only provide (literally the CMB provides ${T_{\rm CMB}=2.725\,{\rm K}}$) $250\,{\rm W}$ in the recent Universe.
If there is a SMBH in the early Universe ($z\sim109$) with a habitable temperature ${T_{\rm CMB}=300\,{\rm K}}$ \citep{Loeb2014}, it would provide $\sim3\times10^{20}\,{\rm W}$.
However, compared to a main sequence star (e.g., $1\,{\rm L_{\odot}}$ $=4\times10^{26}\,{\rm W}$), this available energy is much less.

In this paper, we consider and discuss {six} types of energy sources: the CMB, the Hawking radiation, an accretion disk, {Bondi accretion}, a corona, and the relativistic jets
from two types of black holes: a non-rotating black hole (Schwarzschild black hole) and a rotating black hole (Kerr black hole), ranging from micro, stellar-mass, intermediate-mass to SMBH (i.e, $10^{-5}-10^{10}\,{\rm M_\odot}$).
The structure and outline of this paper are as follows:
we discuss the possible energy sources from a black hole in Sec. \ref{Energy source}.
In Sec. \ref{Dyson Sphere}, we describe the possible location, type, efficiency, and detectability of such a Dyson Sphere.
The conclusion of this study is presented in Sec. \ref{Conclusion}.

\section{Energy source}

In this section, we demonstrate {six} energy sources if a Type II civilisation aims to build a Dyson Sphere around a black hole.
In this paper, we utilise the terminology "promising" and "enough" to indicate when the energy is larger than one solar luminosity for a Type II civilisation.
{We mainly discuss three scenarios: two stellar black holes with masses $5{\rm M_{\odot}}$ and $20{\rm M_{\odot}}$, and a SMBH with mass $4\times10^{6}{\rm M_{\odot}}$.
Their physical parameters are organised in Table \ref{tab1}.}

    \begin{table*}
    	\centering
    	\caption{Three scenarios of a Dyson Sphere around a black hole at different efficiency.}
    	\label{tab1}
    	\begin{flushleft}
    	\begin{tabular}{|c|c|c|c|}\hline
    	Scenario & A(stellar mass black hole) & B(stellar mass black hole) & C(SMBH)\\ \hline \relax 
    	Mass (${\rm M_{\odot}}$) & 5 & 20 & $4\times10^{6}$ \\ \relax
    	Eddington ratio (${\frac{L_{\rm disk}}{L_{\rm Edd}}}$) &$10^{-2}$ & $10^{-1}$ & $10^{-3}$\\ \relax
    	Schwarzschild radius (${\rm km}$) & 14.77 & 59.08 & $1.18\times10^{7}$\\ \relax
    	Dyson sphere radius at $T=3000{\rm K}$ ($R_{\rm sch}$) & {$7.12\times10^{6}$} & {$1.13\times10^{7}$} & {$2.54\times10^{3}$}\\ \relax
        disk luminosity{$^{*}$}   & $1.6\times10^{3}$ (2.38) &$6.4\times10^{4}$ (2.54) &$1.3\times10^{8}$ (2.87)\\ \relax
    	total luminosity{$^{*}$}   & $3.5\times10^{3}$ (2.41) &$1.4\times10^{5}$ (2.57) &$2.8\times10^8$ (2.90)\\ \relax
    	total power{$^{*}$}   & $1.7\times10^{4}$ (2.48) & $6.1\times10^{5}$ (2.63) &$1.1\times10^9$ (2.96)\\   
    	
    	\hline    	\multicolumn{4}{|c|}{50\% efficiency}  \\ \hline \relax
        disk luminosity{$^{*}$}   &{$800$ (2.35)} & $3.2\times10^4$ (2.51) & $6.4\times10^7$ (2.84) \\ \relax
    	total luminosity{$^{*}$}   & {$1750$ (2.38)} & $6.9\times10^4$ (2.54) &$1.4\times10^8$ (2.87) \\ \relax
    	total power{$^{*}$}   & {8450 (2.45)} & $3.1\times10^5$ (2.61) &$5.5\times10^8$ (2.93)\\

    	\hline    	\multicolumn{4}{|c|}{10\% efficiency}  \\ \hline \relax
        disk luminosity{$^{*}$}   & {$160$ (2.28)} & $6.4\times10^3$ (2.44) & $1.3\times10^7$ (2.77) \\ \relax
    	total luminosity{$^{*}$}   & {$350$ (2.31)} & $1.4\times10^4$ (2.47) &$2.8\times10^7$ (2.80) \\ \relax
    	total power{$^{*}$}   & {1690 (2.38)} & $6.1\times10^4$ (2.54) &$1.1\times10^8$ (2.86)\\

    	\hline    	\multicolumn{4}{|c|}{1\% efficiency}  \\ \hline \relax
        disk luminosity{$^{*}$}   & {$16$ (2.18)} & $6.4\times10^2$ (2.34) & $1.3\times10^6$ (2.67) \\ \relax
    	total luminosity{$^{*}$}   & {$ 35 $ (2.21)} & $1.4\times10^3$ (2.37) &$2.8\times10^6$ (2.70) \\ \relax
    	total power{$^{*}$}   & {169 (2.28)} &$6.1\times10^3$ (2.44) &$1.1\times10^7$ (2.76)\\

        \hline
        \end{tabular}\\
        
        {$^{*}$ Numbers outside parentheses are in units of $L_{\odot}$, while numbers inside the parentheses are in units of Kardashev index.}
        \end{flushleft}
    \end{table*}

\label{Energy source}
\subsection{Cosmic Microwave Background}

The CMB is the remnant temperature from the early Universe which is an evidence of the Big Bang.
In the mid-20th century, \citet{Gamow1948} predicted the existence of the CMB while its first measurement was conducted later by \citet{Penzias1965}.
Here, we show the possibility for a Dyson Sphere to collect the CMB energy.

In principle, the CMB is hotter than a black hole (discussed in Sec. \ref{Hawking radiation}).
\citet{Opatrny2017} discussed the concept that a Dyson Sphere absorbs energy from the hot CMB and emits waste energy into a cold black hole.
\citet{Opatrny2017} derived the limiting power from the CMB for a Dyson Sphere as:
\begin{equation}
    \label{limiting energy}
    \begin{split}
    P_{\rm{max}}\sim\frac{27}{256}\sigma ST_{\rm CMB}^{4}, \rm{where} \:\it
    S={\rm 4}\pi \left(\frac{\sqrt{\rm 27}}{\rm 2}R_{\rm Sch}\right)^{\rm 2}=\frac{{\rm 108}\pi G^{\rm 2}M^{\rm 2}}{c^{\rm 4}},
    \end{split}
\end{equation}
where $\sigma$ is the Stefan-Boltzmann constant.
$S$ indicates the surface area of the black hole while $T_{\rm CMB}$ is the temperature of CMB.
$R_{\rm Sch}$ stands for the Schwarzschild radius, $G$ is the gravitational constant, $M$ indicates the mass of the black hole, and $c$ is the light speed in the vacuum.

{{According to \citet{Loeb2014}, the habitable epoch of the Universe starts around when the temperature of the CMB is} ${T_{\rm CMB}=300\,{\rm K}}$. For this purpose, we also discuss the early Universe when the CMB temperature is ${\rm 300\,{\rm K}}$.
The following two conditions, recent Universe with $T_{\rm CMB}=2.725{\rm K}$ \citep{Fixsen2009} and habitable Universe with $T_{\rm CMB}=300{\rm K}$ of $P_{\rm max}$ can be written in the following numerical form, respectively:
\begin{equation}
    \label{recent}
    \frac{P_{\rm{max}}(2.725{\rm K})}{\rm Watt}\sim2.45\times10^{2}\left(\frac{M}{M_{\odot}}\right)^{2},
\end{equation}
and
\begin{equation}
    \label{early}
    \frac{P_{\rm{max}}(300{\rm K)}}{\rm Watt}\sim3.59\times10^{10}\left(\frac{M}{M_{\odot}}\right)^{2}.
\end{equation}}
{We caution readers that black holes when $T_{\rm CMB}=300{\rm K}$ ($z\sim109$) were probably primordial black holes since the first stars are thought to have formed afterwards \citep[e.g.,][]{Carr1977}.}
If a Type II civilisation wants to harvest the CMB power {, no matter in the recent or early Universe, the power is too low to maintain a Type II civilisation. }
Even in the early stage of the Universe, we need to find a SMBH {($M>10^{8}\,{\rm M_{\odot}}$)} to gain the energy of a solar luminosity, which seems to be inefficient.

\subsection{Hawking radiation}
\label{Hawking radiation}

\citet{Hawking1974} predicted and derived the theory of black body radiation emitted from outside the event horizon of a black hole, which was named "Hawking radiation".
Although there is no direct detection hitherto, it is a result of the quantum effects and expected to reduce the black hole mass.

Based on \citet{Hawking1974}, the temperature of a Schwarzschild black hole can be expressed as:
\begin{equation}
    \label{Hawking temperature}
    T_{\rm Sch}=\frac{\hbar c^3}{8 \pi k_{\rm B} GM},
\end{equation}
where $\hbar$ is the reduced Planck constant, and $k_{\rm B}$ stands for the Boltzmann constant.
Nevertheless, we calculate the power of black body luminosity ($L_{\rm Hawking}=4\pi R_{\rm Sch}^{2}\sigma T^{4}_{\rm Sch}$) of Hawking radiation as follows \citep[e.g.,][]{Lopresto2003}:
\begin{equation}
    \label{Hawking luminosity}
    L_{\rm{Hawking}}=\frac{\hbar c^6}{15360 \pi G^2M^2}{\sim9.00\times10^{-29}\left(\frac{M}{M_{\odot}}\right)^{-2}{\rm Watt}}.
\end{equation}
{The} total radiation from a black hole itself is extremely low.

In terms of a rotating black hole, the temperature of the Hawking radiation from a rotating black hole (here we assume Kerr-Vaidya metrics) can be expressed as follows \citep[e.g.,][]{Chou2020}:
\begin{equation}
    \label{eqHawking radiation}
    T_{\rm Kerr}=\gamma\frac{\hbar c^3}{8 \pi k_{\rm B} GM},
    \rm{where} \: \it
    \gamma=\frac{\sqrt{\rm 1-{\it a}^2}}{\rm 1+\sqrt{1-{\it a}^2}},
\end{equation}
where $a$ stands for the spin parameter of a black hole.
As the spin parameter increases, the factor $\gamma$ decreases.  
We conclude that collecting the Hawking radiation from a Kerr black hole is harder than a Schwarzschild black hole (although both ways are almost impossible).
We will not discuss the luminosity of a Kerr black hole further since it is harder to harvest the energy from a Kerr black hole.

    

\subsection{Accretion disk}
The massive object in the centre would accrete the gas and the material close to it into a vortex disk.
Due to the strong gravity from a black hole, the material in an accretion disk would be heated by friction.
After the igniting process, an accretion disk emits strong radiation.
Hence, a Type II civilisation could harvest the radiation from the accretion disk.

\begin{figure}
    	\includegraphics[width=\columnwidth]{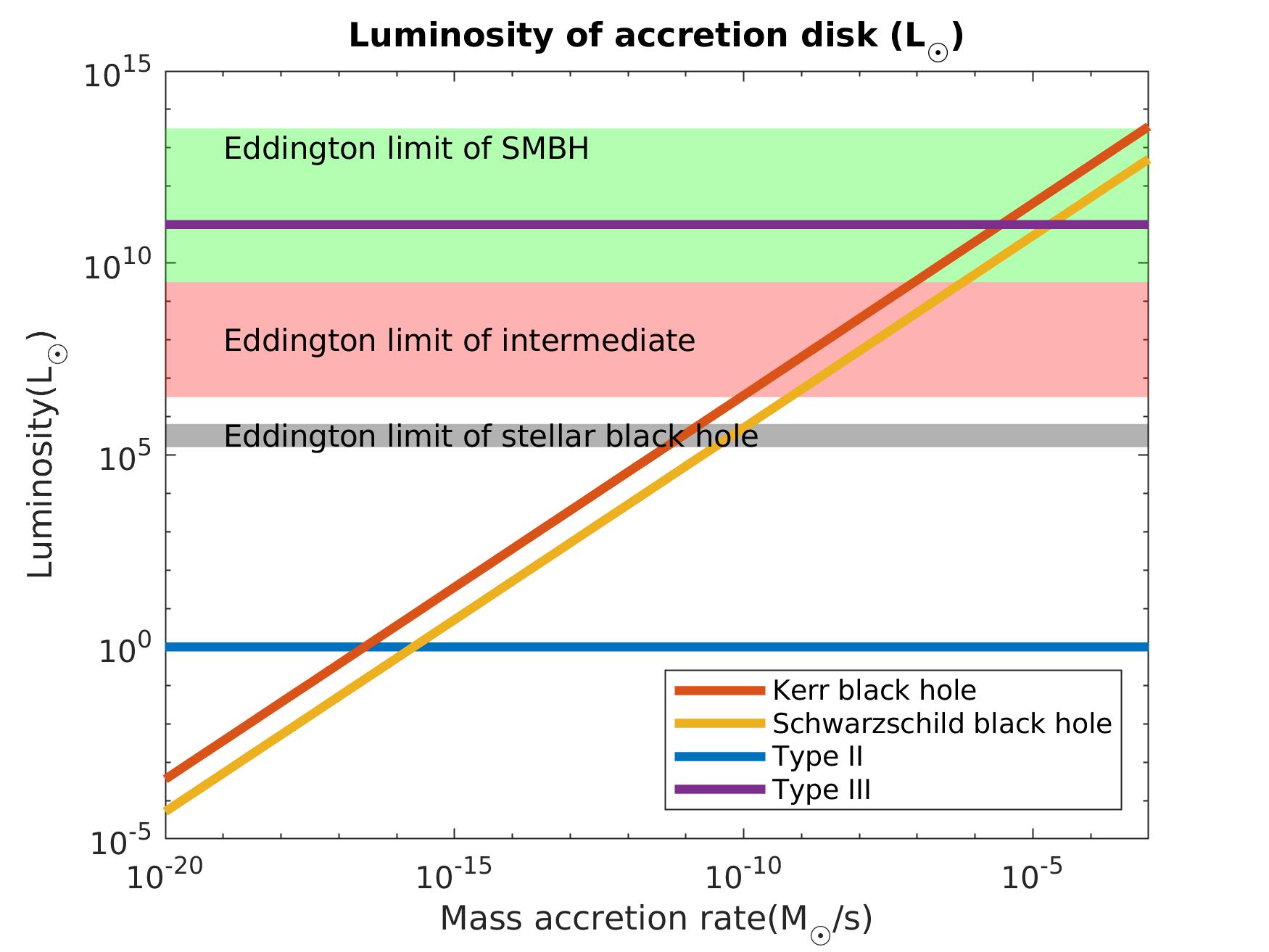}
    	\caption{Accretion disk luminosity as a function of the mass accretion rate. The orange line suggests $\eta=0.399$ which is the case of the extreme Kerr black hole ($a=1$) while the yellow line shows $\eta=0.057$ which is the case of non-rotating black holes. The blue line and the purple line indicate the required energy of a Type II and a Type III civilisations, respectively. Shaded regions suggest Eddington luminosity limits of different types of black hole: grey for stellar-mass black holes ($5\,{\rm M_{\odot}}<M<20\,{\rm M_{\odot}}$), red for intermediate-mass black holes ($10^2\,{\rm M_{\odot}}<M<10^5\,{\rm M_{\odot}}$) and green for SMBHs ($10^5\,{\rm M_{\odot}}<M<10^9\,{\rm M_{\odot}}$).}
    	
        \label{figdisk}
    \end{figure}

We first consider the Eddington luminosity of a black hole in different mass ranges \citep{Rybicki1979}:
\begin{equation}
    \label{Eddington luminosity}
    L_{\rm{Edd}}=\frac{4\pi GMm_{\rm p}c}{\sigma_{\rm T}}{\sim3.2\times10^{4}\left(\frac{M}{M_{\odot}}\right)L_{\odot}},
\end{equation}
where $m_{\rm p}$ suggests the mass of a proton, and $\sigma_{\rm T}$ is Thomson scattering cross-section ($\sigma_{\rm T}=6.65\times10^{-29}\,{\rm m^{-2}}$).
{The result suggests} that if a Type II civilisation finds a black hole with accretion disk reaching the Eddington limit, even if the black hole is only the size of a stellar black hole ($5\,{\rm M_{\odot}}<M<20\,{\rm M_{\odot}}$), it could provide up to $\sim10^{5}\,{\rm L_{\odot}}$.
Additionally, we consider the luminosity as a function of mass accretion rate:
\begin{equation}
    \label{Disk luminosity}
    L_{\rm{disk}}=\eta_{\rm disk}{\frac{dm}{dt}}c^2,
\end{equation}
where $\eta_{\rm disk}$ is the accretion efficiency, $dm/dt$ indicates the mass accretion rate.
According to \citet{Thorne1974}, the accretion efficiencies of different types of black holes are $\eta_{\rm disk}=0.057$ for a Schwarzschild black hole and $\eta_{\rm disk}=0.399$ for an Extreme Kerr black hole ($a=1$), respectively.
Adopting these efficiencies, we calculate the luminosity at different accretion rates.
The result is shown in Fig. \ref{figdisk}.
The different coloured regions indicate the Eddington limit of stellar-mass black holes ($5\,{\rm M_{\odot}}<M<20\,{\rm M_{\odot}}$), intermediate-mass black holes ($10^2\,{\rm M_{\odot}}<M<10^5\,{\rm M_{\odot}}$), and SMBHs ($10^5\,{\rm M_{\odot}}<M<10^9\,{\rm M_{\odot}}$).
The results suggest that even for a stellar-mass black hole with a low Eddington ratio ($10^{-3}$), the luminosity of the accretion disk is several hundred times of a solar luminosity.

Furthermore, an intermediate-mass black hole with a high accretion rate would emit an energy up to $\sim 10^{9}\,{\rm L_{\odot}}$, which is only 1\% of the required energy of a Type III civilisation. 
At the same mass accretion rate, the luminosity of an extremely Kerr black hole accretion disk is seven times larger than the Schwarzschild black hole due to the accretion efficiency.
Therefore, we conclude that the accretion disks of both types of black holes (non-rotating and extremely rotating) have the potential as a Type II civilisation's energy source.

    
{
\subsection{Bondi Accretion}
Bondi accretion is a spherical gas accretion onto a compact gravitating object from the interstellar medium with no angular momentum and no magnetic field \citep{Hoyle1939,Bondi1952}.
Considering the Bondi accretion around the vicinity of black holes, the analytical solution of the luminosity from Bondi accretion can be expressed as follows:
\begin{equation}
L_{\rm Bondi}=\eta c^{2}\left(\frac{\pi G^{2} M^{2} \rho}{c_{s}^{3}}\right),
\end{equation}
where $L_{\rm Bondi}$ is the luminosity of Bondi accretion, $\rho$ is the ambient density, and $c_{s}$ is the sound speed.
The sound speed in a gas cloud can be expressed as follows:
\begin{equation}
c_{s}=\sqrt{\frac{k_{\rm B}T_{\rm gas}}{\mu m_{\rm H}}},
\end{equation}
where $T_{\rm gas}$ is the temperature of the gas, $\mu$ is molecular weight, and $m_{\rm H}$ is mass of a hydrogen atom.
Adopting $\mu=1$, $m_{\rm H}=1.67\times10^{-27}\,{\rm kg}$, $T_{\rm gas}=10^{7}\,{\rm K}$ for the SMBH \citep[e.g.,][]{DiMatteo2003} and $T_{\rm gas}=10^{4}\,{\rm K}$ for the stellar mass black holes \citep[e.g.,][]{Alvarez2009}, and $\rho=1.67\times10^{-27}\,{\rm kg/cm^{3}}$, the Bondi luminosity can be expressed as:
{
\begin{equation}
L_{\rm Bondi}\sim2.8\times10^{3}\left(\frac{M}{M_{\odot}}\right)^{2}\left(\frac{n}{\#/{\rm cm}^{3}}\right)\left(\frac{T}{\rm K}\right)^{-3/2}.
\end{equation}
}
For scenarios A, B, and C, we obtained $0.07$, $1.11$, $1.40\times10^{6}\,L_{\odot}$, respectively.
These are $\sim10^{-6}$ and $\sim1\%$ of their accretion disk luminosities for the stellar mass black holes and the SMBH, respectively.
These radiative energies are too faint compared to the energy from the accretion disk.}

\subsection{Corona}
The gas around a black hole is in the plasma state due to the high temperature from the accretion disk, and this surrounding gas is called the corona.
Meanwhile, the corona generates high energy radiation as well \citep[e.g.,][]{Haardt1993}.
Thus, the radiation from the corona could be a good candidate in providing additional energy, aside from the radiation from the accretion disk.

For simplicity, we adopt the empirical relation between the corona luminosity and the disk luminosity from \citet{Sazonov2012}:
\begin{equation}
    \label{corona}
    {\rm log}\left(\frac{L_{\rm corona}}{\rm 10^{44}\,{\rm erg\ s^{-1}}}\right)={\rm 1.03log}\left(\frac{L_{\rm disk}}{\rm 10^{44}\,{\rm erg\ s^{-1}}}\right)-0.207,
\end{equation}
where $L_{\rm corona}$ represents the luminosity of corona in the unit $10^{44}\,{\rm erg\ s^{-1}}$.
This relation is the best-fit of the luminosity of hot corona versus disk luminosity using 60 active galactic nuclei from International Gamma-Ray Astrophysics Laboratory (INTEGRAL) all-sky hard X-ray survey.
{The samples have similar X-ray flux variations with X-ray binaries, which can be expressed as a log-normal distribution.
We assume that this formula extends to different mass ranges including stellar mass black holes \citep[e.g.,][]{Uttley2005}.}
For a stellar-mass black hole, the corona luminosity is approximately one-third of the radiation from the accretion disk while the ratio between the corona luminosity and the accretion disk luminosity increases to $1/2$ for a SMBH.
If a Type II civilisation includes the corona luminosity as well, the useful energy becomes 1.3 to 1.5 times of the accretion disk luminosity, which would be another promising energy source for maintaining civilisation.


\subsection{Relativistic jets}

Relativistic jets are high-energy and piercing phenomena, which are common among compact objects.
These plasma flow contain tremendous energy and are usually along the axis of rotation of the object.
The particles inside the jets could reach the Lorentz factor up to $100$ ($\gamma=(\sqrt{1-v^2/c^2})^{-1}$).
Although the relation between the accretion disk and the relativistic jets is not clear so far, it is believed that the jets are driven by the tangled magnetic field \citep[e.g.,][]{Blandford1977,Hawley2002,McKinney2004}.

For simplicity, according to \citet{Ghisellini2014}, the radiation of relativistic jets has an empirical relation with the luminosity of the accretion disk for a SMBH as follows:

\begin{equation}
    \label{jetsdisk}
    {\rm log}\left(\frac{L_{\rm jets}}{\rm erg\ s^{-1}}\right)={\rm 0.98log}\left(\frac{L_{\rm disk}}{\rm erg\ s^{-1}}\right)+0.639,
\end{equation}
where ${L_{\rm jets}}$ is the electromagnetic radiation of the jets. 
This relation is based on the least-squares best fit of the radiative jet power versus disk luminosity using objects detected with Fermi/Large Area Telescope (LAT) and spectroscopic optical observations.
{Most of the Eddington ratios of the samples are in the range between $1\%$ to $100\%$, suggesting that the disks are geometrically thin, \citep[e.g.,][]{Shakura1973}.
X-ray binaries in the soft state are usually in this case \citep[e.g.,][]{Fender2002}.
Therefore, we assume that this empirical formula holds in different mass ranges.}
A stellar-mass black hole with a low Eddington ratio can even produce several orders of solar luminosity, suggesting that the energy in jets has the potential to provide the energy required by a Type II civilisation.

According to Eq. \ref{jetsdisk}, the radiation from the jets would be $\sim80\%$ and $\sim60\%$ of the radiation from an accretion disk for a stellar-mass black hole and a SMBH, respectively.
In addition, the radiation from the jets is believed to be $\sim10\%$ of the total energy of jets \citep[e.g.,][]{Ghisellini2014}.
If the technology is able to collect the energy from not only the luminosity but also the total energy of the jets, the energy available would be 10 times larger.
According to our results, only with the mass $M=0.003\,{\rm M_{\odot}}$ and Eddington ratio$=10^{-3}$, could the energy reach one solar luminosity.
For a stellar-mass black hole with $5\,{\rm M_{\odot}}$, the total jet energy becomes $\sim1.7\times10^{6}\,{\rm L_{\odot}}$. 
The results suggest that the relativistic jets could be promising sources in the Universe regardless of the high temperature.


\section{Dyson Sphere}
\label{Dyson Sphere}
In this study, we only discuss the civilisations which were born and raised from other stars.
We speculate that this kind of civilisation can collect the energy remotely or treat the energy source as a power station rather than living around a black hole with a harsh environment.
Therefore, throughout the paper, we do not discuss whether the temperature and the gravity of our configurations are suitable for life \citep[e.g.,][]{Schwartzman1977,Inoue2011}.
{Moreover, we discuss two scenarios for the Dyson Sphere: (1) assuming that a Type II civilisation has the technology advanced enough to afford extreme high temperatures (hot Dyson Sphere hereafter) and (2) the realistic solid material to avoid melting $T\sim3000\,{\rm K}$ (solid Dyson Sphere hereafter)\footnote[6]{\url{https://pubchem.ncbi.nlm.nih.gov/compound/23964}}.} 

\subsection{Possible Type and Location}
In reality, a monolithic Dyson Sphere is mechanically unstable due to the pressure and the gravity from the central star \citep[e.g.,][]{Wright2020}.
Hence, a structure similar to a Dyson Swarm or a Dyson Bubble becomes a more promising candidate to collect energy.
Moreover, considering a Type II civilisation might want to transfer energy from jets as well, a Dyson Bubble would be a more possible case to collect the energy from jets since the celestial mechanics of a Dyson Swarm is complicated and chaotic.
Therefore, we discuss two scenarios for a Dyson Sphere.
At the early stage of building a Dyson Sphere, the most simplified configuration is a Dyson Ring (a type of a Dyson Swarm), which contains numerous collectors sharing the same orbit.

Taking the widely discussed stellar-mass black hole, Cygnus X-1, as an example \citep[e.g.,][]{Shapiro1976,Ichimaru1977,Fabian1989}, with the mass of $M=8.7\,{\rm M_{\odot}}$ \citep{Iorio2008}, the accretion disk extends to ${r\sim500R_{\rm Sch}}$ \citep{Young2001}.
For a simplified discussion, we assume the inner radius of the accretion disk is around ${R_{\rm in}=3R_{\rm Sch}}$, which is the radius outside the {innermost stable circular orbit (ISCO)} of a Schwarzschild black hole.
In terms of the outer radius of the accretion disk, which has not been studied well so far, we assume its outer radius extends to ${R_{\rm out}=1000R_{\rm Sch}}$ \citep[e.g.,][]{Inoue2011,You2012}.
For a stellar-mass black hole, since its accretion disk size is smaller than the size of the Sun, the radius of its Dyson Sphere is smaller than that of a stellar Dyson Sphere (${R_{\rm out}=1\,{\rm AU}}$ corresponds to $M=5\times10^{4}\,{\rm M_{\odot}}$).
At this stage, the possible radius for the Dyson Sphere to collect the energy from the accretion disk should be around $10^{3}R_{\rm Sch}<R_{\rm DS}<10^{5}R_{\rm Sch}$, which locates outside the accretion disk.
If a Dyson Sphere is constructed farther away from the accretion disk, the efficiency may be too low (e.g., if two Dyson Spheres of the same size are built at distances of $10^{3}R_{\rm Sch}$ and $10^{6}R_{\rm Sch}$, respectively, the latter one could only intercept $10^{-6}$ times the energy of the former one), and might crush into the dusty torus.

{
For a given black hole mass, luminosity, and temperature for a Dyson Sphere to afford, we can rewrite the luminosity of a spherical blackbody ($L=4\pi \sigma R^{2}T^{4}$) into the following form to calculate the radius of a Dyson Sphere in units of $R_{\rm Sch}$:
\begin{equation}
    \label{DSradius}
\left(\frac{R_{\rm DS}}{R_{\rm Sch}}\right)=\frac{c^{2}}{2GM}\sqrt{\frac{L_{\rm disk}}{4\pi \sigma T^4}}\sim8.01\times10^{12}\left(\frac{M}{M_{\odot}}\right)^{-1}\left(\frac{L}{L_{\odot}}\right)^{\frac{1}{2}}\left(\frac{T}{{\rm K}}\right)^{-2},
    \end{equation}
where $T$ is the temperature a Dyson Sphere wants to absorb rather than emitting.
In terms of the solid material which can absorb $\sim3000K$, the Dyson Sphere should be located at $R_{\rm DS}\gtrsim7.12\times10^{6}R_{\rm Sch}$ and $R_{\rm DS}\gtrsim1.13\times10^{7}R_{\rm Sch}$ for Scenario A and B, respectively, to avoid melting.
For a SMBH, Scenario C, the Dyson Sphere could be set at $R_{\rm DS}\gtrsim2.54\times10^{3}R_{\rm Sch}$.}

{Not surprisingly, if a Type II civilisation wishes to build a Dyson Sphere receiving $T=3000\,{\rm K}$ and $1\,{\rm L_{\odot}}$, the total area required is independent from the mass of the source black hole (or a star), which is $\sim8.65\times10^{19}\,{\rm m^{2}}$.
Compared to the total area ($7.03\times10^{22}\,{\rm m^{2}}$) of the Dyson Sphere around the Sun at $1\,{\rm AU}$, this area is $\sim100$ times smaller.
Indeed, if the Dyson Sphere is closer to the black hole, the temperature is higher but the needed surface area decreases.
However, the maximum luminosity a Type II civilisation can absorb is different.}

In terms of energy from jets, utilising the pressure that balances the inward gravity and without the need to be in orbit (stationary in space), a Dyson Bubble is considered to be built with a huge amount of light sails.
First of all, we consider the balance between different types of pressures and the gravitational force from a black hole:
\begin{equation}
    \label{lightsail}
F_{\rm G}=\oint_S {(p_{\rm rad}+p_{\rm dyn}+p_{\rm B})} \,ds,
    \end{equation}
where $S$ is the surface area of a light sail, and the notations of $p$ stands for different types of pressure outwards:
radiation pressure ($p_{\rm rad}\propto I$ {$\propto R_{\rm DS}^{-2}$}; $I$: flux (${\rm W/m^{2}}$)), dynamical pressure ($p_{\rm dyn}\propto\rho v^{2}$ {$\propto R_{\rm DS}^{-3}$}; $\rho$: density of the jets and $v$: speed of jets) and magnetic pressure ($p_{\rm B}\propto B^{2}$ {$\propto R_{\rm DS}^{-4}$}; $B$: magnetic flux density).
{Note that $R_{\rm DS}$ stands for the distance of the Dyson Sphere to the black hole rather than the radius of the light sail itself.}
Along the path of jets, pressure decreases as radius increases {($p_{\rm{total}}\propto \frac{c_{1}}{R^{2}_{\rm DS}}+\frac{c_{2}}{R^{3}_{\rm DS}}+\frac{c_{3}}{R^{4}_{\rm DS}}$,
where
$c_{1}$, $c_{2}$, and $c_{3}$ represent the coefficients of different types of pressures).}
Moreover, the increase of the surface area of a light sail increases the outward force on the sail.
The surface area of a light sail is proportional to $R^{2}_{\rm DS}$ if a light sail is designed to cover the whole opening angle of the jet{, which leads to $F_{\rm G}\propto \frac{m}{R^{2}_{\rm DS}} \propto R^{2}_{\rm DS}\left(\frac{c_{1}}{R^{2}_{\rm DS}}+\frac{c_{2}}{R^{3}_{\rm DS}}+\frac{c_{3}}{R^{4}_{\rm DS}}\right)$}.
Considering an outward pressure and inward gravitational force, the mass $m$ required to build a light sail would be smaller when a light sail is closer to the black hole {($m\propto c_{1}R^{2}_{\rm DS}+c_{2}R^{1}_{\rm DS}+c_{3}$)}.
Thus, to save the materials, it is better to locate a light sail close to where the jets originate from {no matter how large the coefficients are}.

\begin{figure*}
    	\includegraphics[width=\textwidth ]{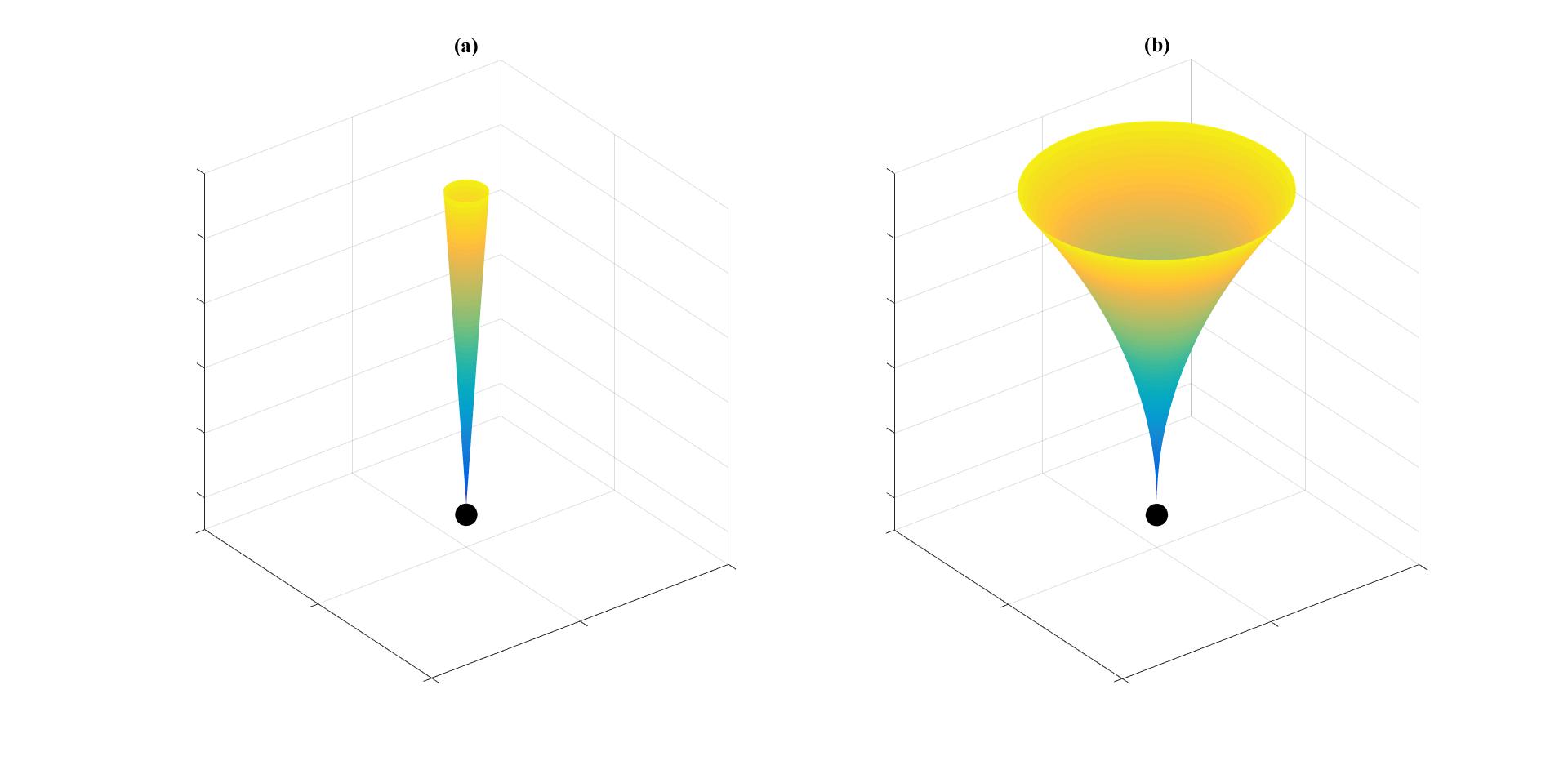}
    	\caption{Schematic of the relativistic jets. (a) linear transmission of jets (i.e., the surface area $\propto R^{2}_{\rm DS}$) and (b) with changing angle (i.e., the surface area $\propto R^{k}_{\rm DS}$, here $k=4$).
    	The color scales with the distance from the black hole. Scales are arbitrary. The panels are not to scale.}
        \label{figscheme}
    \end{figure*}

{If} the jet opening angle also changes with the radius, we assume $S\propto R^{k}_{\rm DS}$, where $S$ is the surface area of a light sail and $k$ is an arbitrary real number.
Eq. \ref{lightsail} becomes {$\frac{m}{R^{2}_{\rm DS}}\propto R_{\rm DS}^k( \frac{c_{1}}{R^{k}_{\rm DS}}+\frac{c_{2}}{R^{k+1}_{\rm DS}}+\frac{c_{3}}{R^{4}_{\rm DS}})$.}
Regardless of the value of $k$, the mass of a light sail required decreases at the location closer to the origin of the jets.
These two scenarios are visualised in Fig. \ref{figscheme}.
{Fig. \ref{figscheme}(b) shows the schematic of changing angle jets  described by an arbitrary power law.}
Under these assumptions, a Type II civilisation could build a light sail with fewer materials.
However, a Type II civilisation needs to consider the ISCO ($r={\frac{6GM}{c^{2}}}$ and $r={\frac{GM}{c^{2}}}$ for a Schwarzschild black hole and an extreme Kerr black hole, respectively) if a Type II civilisation wants to collect energy in situ instead of remotely.
{However, we caution readers that the luminosity of accretion disks of black holes varies with time.
Accretion disks switch between bright and dim states \citep[e.g.,][]{Done2007}.
If the accretion disk becomes too bright, the structures might be blown away while if the accretion disk gets too faint, the Dyson Sphere may fall into the black hole.
}

\subsection{Efficiency}
Even in reality, considering the transfer efficiency and the covering fraction, the accretion disk could still play a pivotal role for a Type II civilisation.
We combine two concepts, energy transfer efficiency and the sky covering fraction, into the terminology "efficiency" (e.g., if a Dyson Sphere covers 10\% of the sky and 50\% of energy transfer to usable energy, the efficiency is 5\%).
Most of the solar cell efficiency on the Earth reaches $15\%$.
In the world record, the highest efficiency is $\sim47\%$ \citep{Geisz2018}.
For instance, if there is a 50\% energy-conversion efficiency in the future, a Dyson sphere covering 2\% of the sky can extract 1\% available energy from the 5 M$_{\odot}$ black hole.
This corresponds to {16 $L_{\odot}$} which is sufficient for a Type II civilisation.
In this case, only 10\% surface area of the Earth is needed to cover the 2\% of the sky at a distance of $r_{\rm out}=1000R_{\rm Sch}$.

Assuming a Type II civilisation collects the energy from a $5\,{\rm M_{\odot}}$ black hole with a low Eddington ratio {(${\frac{L_{\rm disk}}{L_{\rm Edd}}}=10^{-2}$)}, the total luminosity from the accretion disk is {$1600\,{\rm L_{\odot}}$} (${\frac{L_{\rm disk}}{L_{\rm Edd}}}\times L_{\rm Edd}$).
Moreover, if a Type II civilisation could find a stellar-mass black hole with higher Eddington ratio (${\frac{L_{\rm disk}}{L_{\rm Edd}}}=10^{-1}$) and larger mass ($M=20\,{\rm M_{\odot}}$), the total available energy from the accretion disk is $6.4\times10^{4}\,{\rm L_{\odot}}$.
In other words, the accretion disk alone can provide sufficiently enormous luminosity to be collected by a Type II civilisation.
If a Type II civilisation could find such a way, it could utilise an accretion disk rather than hunting ten thousand stars.
The black hole could play an important role for a Type II civilisation whether or not they just reach a Type II civilisation or will soon become a Type III civilisation.
Furthermore, if a Type II civilisation becomes a Type III civilisation in a short period, it will likely seek a SMBH located at the galactic centre such as Sgr A* in the Milky Way.
If a SMBH with the mass ($M=4\times10^{6}\,{\rm M_{\odot}}$) and low Eddington ratio (${\frac{L_{\rm disk}}{L_{\rm Edd}}}=10^{-3}$), the total luminosity from the accretion disk is $10^{8}\,{\rm L_{\odot}}, \rm{or} \: 10^{34}\,{\rm W}$, which makes up $0.025\%$ of the required by a Type III civilisation ($4\times10^{37}{\,\rm W}$).
It would be much more efficient than collecting a host of stars distributed in a galaxy.

Additionally, if a Type II civilisation not only absorbs the energy from an accretion disk but also collects the radiation from a corona and relativistic jets, the available energy would be larger.
We consider two kinds of energy: total luminosity (${L_{\rm total}=L_{\rm disk}+L_{\rm corona}+L_{\rm jets}}$) and total energy including the kinetic energy of the jet (${P_{\rm total}=P_{\rm jets}+L_{\rm total}}$).
For the first scenario ($5\,{\rm M_{\odot}}$ and {${\frac{L_{\rm disk}}{L_{\rm Edd}}}=10^{-2}$}), if 1\% energy could be transferred to usable energy, the usable energy are ${L_{\rm total}\sim{35}\,{\rm L_{\odot}}}$ and ${P_{\rm total}\sim{169}\,{\rm L_{\odot}}}$ while Kardashev indices could reach {$K=2.21$ and $K=2.28$}.
As for the second scenario ($20\,{\rm M_{\odot}}$ and ${\frac{L_{\rm disk}}{L_{\rm Edd}}}=10^{-1}$), the total luminosity and the total power are $1.4\times10^5\,{\rm L_{\odot}}$ and $6.1\times10^5\,{\rm L_{\odot}}$, which could boost the civilisation up to $K=2.57$ and $K=2.63$.
In the last scenario for the SMBH ($4\times10^{6}\,{\rm M_{\odot}}$ and ${\frac{L_{\rm disk}}{L_{\rm Edd}}}=10^{-3}$), the total power is $\sim10^{9}\,{\rm L_{\odot}}$, giving $K\sim3$.
We organise different configurations of power for three scenarios with different efficiencies in Table \ref{tab1}.
    
\subsection{Detectablility}
In terms of detectability, direct evidence from a Dyson Sphere seems to be impossible because even if a Type II civilisation covers 90\% of the luminosity from the accretion disk, we can only expect an Eddington ratio 10 times smaller.
For instance, if there is a black hole with the Eddington ratio $\frac{L_{\rm disk}}{L_{\rm Edd}}=10\%$ in the true Universe and a Type II civilisation covers 90\% of the luminosity from the accretion disk, we would falsely measure the Eddington ratio of this black hole as 1\%.

\begin{figure}
    	\includegraphics[width=\columnwidth]{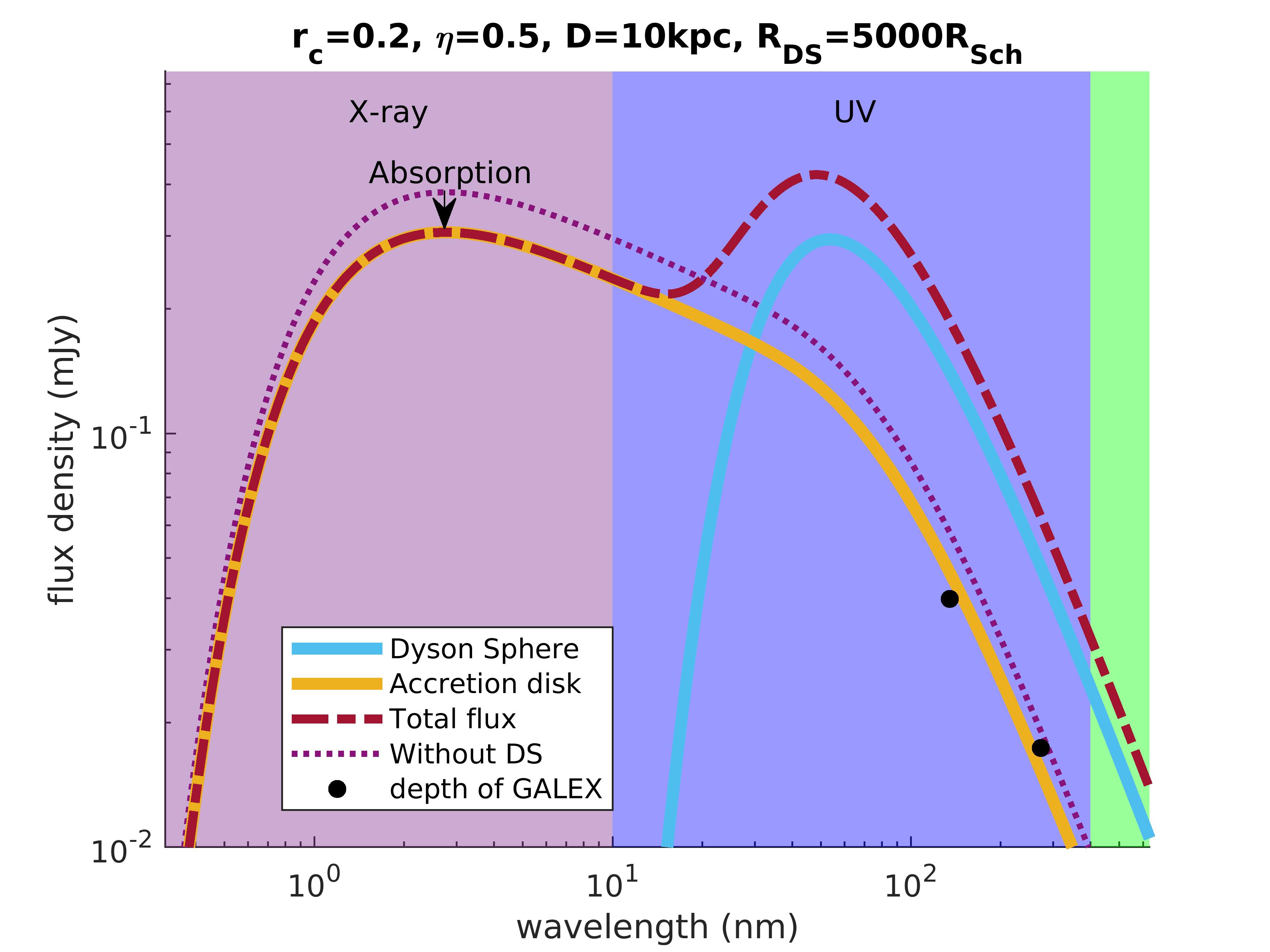}
     	\includegraphics[width=\columnwidth]{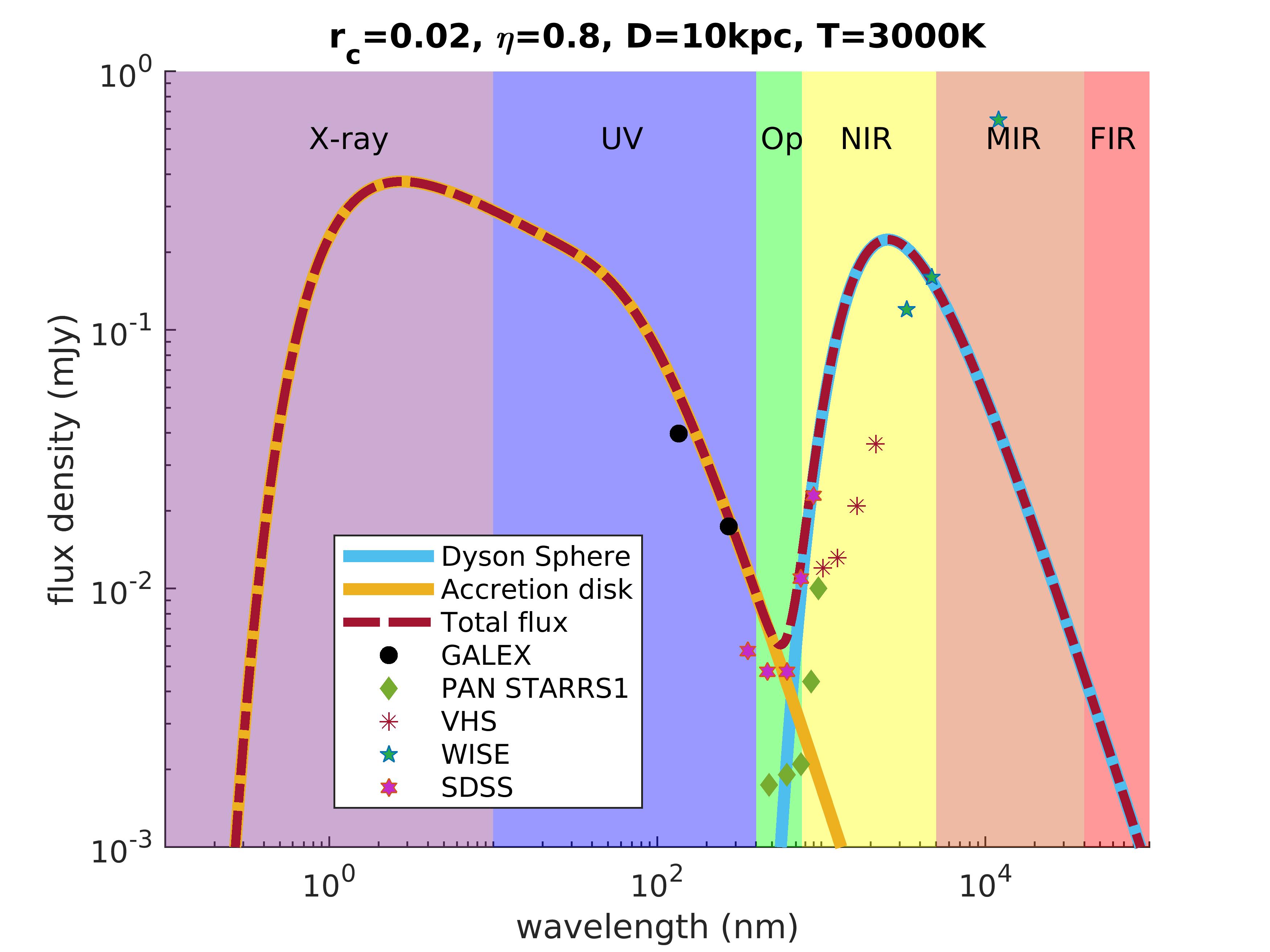}
    	\caption{Example spectra of Scenario A ($M=5\,{\rm M_{\odot}}$ and {$L_{\rm disk}=1600\,{\rm L_{\odot}}$)} {(upper panel:)} A hot Dyson Sphere with covering fraction $R_{\rm c}=20\%$, transfer efficiency $\eta=50\%$ {, and $R_{\rm DS}=5000R_{\rm Sch}$. (lower panel:) A solid Dyson Sphere with covering fraction $R_{\rm c}=2\%$, transfer efficiency $\eta=80\%$, and $R_{\rm DS}=7.12\times10^{6}R_{\rm Sch}$}.
    	The yellow curve suggests the accretion disk flux with the absorption of the Dyson Sphere (($1-R_{\rm c}$) times the original accretion disk flux) while the purple dot curve is the original flux of accretion disk without surrounding a Dyson Sphere. The blue curve shows the flux of the waste heat from the Dyson Sphere while the red dot-line curve indicates the total flux.
    	{The black circles, green diamonds, burgundy asterisks, blue stars, and magenta hexagrams indicate the limiting magnitude of {\it GALEX}, PAN STARRS1, {\it VHS}, {\it WISE}, and {\it SDSS} survey, respectively.}
    	}
        \label{figdet}
    \end{figure}

A more promising way to detect Dyson Spheres is from the waste heat in IR.
After utilising the energy, the wasted IR energy would be released to the background by a typical Dyson Sphere.
Here we calculate the spectrum of an accretion disk with a Dyson Sphere.
We assume multi-colour black body radiation \citep[e.g.,][]{Mitsuda1984,Makishima1986,Merloni2000} for the disk spectrum:
\begin{equation}
    \label{dbb}
f_{\rm disk}=\frac{\rm{cos(\theta)}}{D^2}{\int_{r_{\rm in}}^{r_{\rm out}}}2\it{\pi r{B(T(r))}\,dr},
    \end{equation}
adopting the temperature profile $\propto r^{-3/4}$ (optically thick{, geometrically thin} accretion disk).
{$f_{\rm disk}$ is the flux of the accretion disk. $\theta$ stands for the inclination angle of the disk ($\theta=0^{\circ}$ in our spectrum), $D$ is the distance between the observer and the disk, while {$B(T(r))$} is the Planck function.} 
The luminosity of an accretion disk is as follows:
\begin{equation}
    \label{dbbL}
L_{\rm disk}=\int_{r_{\rm in}}^{r_{\rm out}}4\pi r\sigma T(r)^4\,dr\sim4\pi r_{\rm in}^2\sigma T_{\rm in}^4.
    \end{equation}
We assume Scenario A with the inner radius $r_{\rm in}=3R_{\rm sch}$ and the outer radius $r_{\rm out}=1000R_{\rm sch}$, and according to Eq. \ref{dbbL}, { $T_{\rm in}\sim4.6\times10^6\,{\rm K}$}.
In addition, we assume the transfer efficiency is $\eta=50\%$ and the covering fraction is $R_{\rm c}=50\%$ {for the hot Dyson Sphere while $\eta=80\%$ and $R_{\rm c}=2\%$ for the solid Dyson Sphere.}
The distance between the black hole and the observer is set at $10\,{\rm kpc}$.
The resulting {hot and solid Dyson Sphere} spectra of the accretion disk, waste heat, and total flux which we could measure are shown in Fig. \ref{figdet}.
In {these configurations}, the AB magnitude ($m_{\rm AB}$) of the waste heat at {$10\,{\rm nm}$ to $300\,{\rm nm}$ and $10^{3}\,{\rm nm}$ to $10^{4}\,{\rm nm}$ are} approximately {$17$ to $20\,{\rm mag}$ and $18$ to $20\,{\rm mag}$ for the hot and solid Dyson Sphere, respectively.} 
{The limiting magnitude of Wide Field Camera 3 (WFC3)/UVIS of Hubble Space Telescope ({\it HST}) in NUV(F225W) filter over 1 hour is $\sim27$ mag\footnote[7]{\url{https:// HST-docs.stsci.edu/wfc3ihb/chapter-6-uvis-imaging-with-wfc3/6-8-uvis-sensitivity}}, suggesting that in this scenario, the hot Dyson Sphere could be detected by our current ultraviolet (UV) space telescope.}
With the {\it HST}, one can look for these around known black holes/X-ray binaries because the {\it HST}‘s Field of View (FoV) is not so large.
Alternatively, if one wants to survey a large area, a telescope with a larger FoV would be more suitable such as Galaxy Evolution Explorer ({\it GALEX}) Ultra-violet Sky Surveys, {\it SDSS}, Pan-STARRS1 $3\pi$ Steradian Survey (optical), VISTA Hemisphere Survey ({\it VHS}; near-infrared),Wide-Field Infrared Survey Explorer ({\it WISE}; mid-infrared), and Sloan Digital Sky Survey ({\it SDSS}; optical).
{These limiting magnitudes are summarised in Table \ref{tab2} and included in Fig. \ref{figdet}.}

{
If a black hole is closer to us and has a higher mass, the source will be brighter than our scenario ($D=10{\,\rm kpc}$ and $M=5\,M_{\odot}$).
For instance, if the source is at $1\,{\rm kpc}$, the AB magnitude of the waste heat at $10\,{\rm nm}$ to $300\,{\rm nm}$ and $10^{3}\,{\rm nm}$ to $10^{4}\,{\rm nm}$ are approximately $12$ to $13$ and $13$ to $17\,{\rm mag}$ of the hot Dyson Sphere and the solid Dyson Sphere under the same configurations.
If the source is under the scenario B ($M=20\,M_{\odot}$ and Eddington ratio$=0.1$), the AB magnitude of the waste heat at $10\,{\rm nm}$ to $100\,{\rm nm}$ and $10^{3}\,{\rm nm}$ to $10^{4}\,{\rm nm}$ are approximately $13$ to $15$ and $14$ to $18\,{\rm mag}$ of the hot Dyson Sphere and the solid Dyson Sphere at $10\,{\rm kpc}$.
}

\begin{figure}
    	\includegraphics[width=\columnwidth]{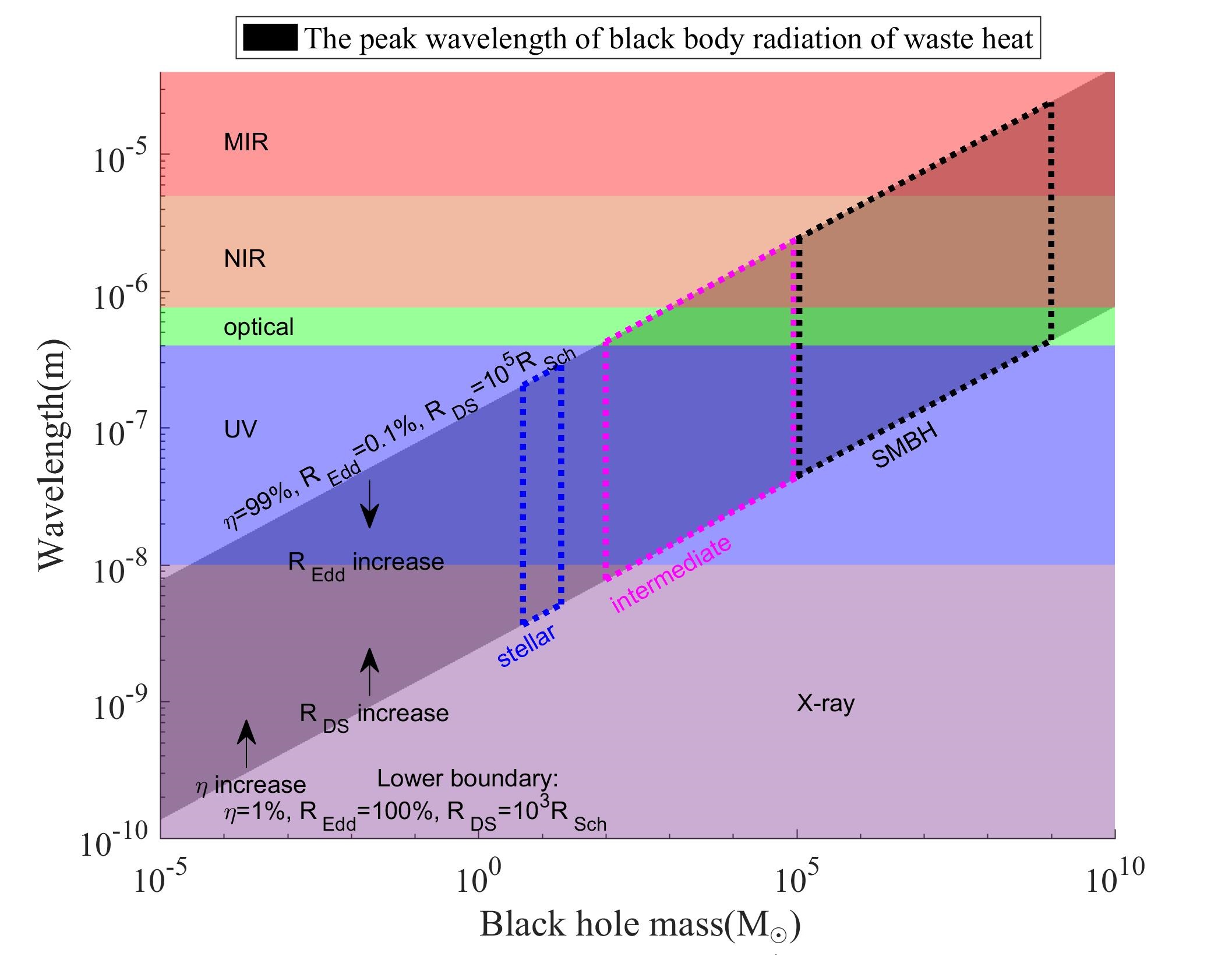}
    	\caption{Possible wavelengths for a {hot} Dyson Sphere to be detected in different black hole mass.
    	The purple, blue, green, orange, and red regions indicate X-ray$(\rm 0.01-10\,{\rm nm})$, UV$(\rm 10-400\,{\rm nm})$, optical$(\rm 400-760\,{\rm nm})$, NIR($\rm 760\,{\rm nm}-5\,{\rm \mu m}$), and MIR($\rm 5-40\,{\rm \mu m}$) wavelength, respectively.
    	The shaded region is the peak wavelength from black body radiation of the waste heat, covering a wide range of parameters of a Dyson Sphere.
    	The region enclosed by blue, magenta, and black lines indicate stellar-mass, intermediate-mass and SMBH.
    	Three arrows show how the peak wavelengths change with increasing parameters.
    	The arrow lengths are arbitrary.}
        \label{figwav}
    \end{figure}

Interestingly, with little algebra ($L_{\rm DS}=R_{\rm c}(1-\eta)L_{\rm disk}$), we derive the temperature of the waste heat, which is not related to the covering fraction:
\begin{equation}
    \label{wasteT}
T_{\rm waste}=\left[\frac{(1-\eta)L_{\rm disk}}{4\pi \sigma R_{\rm DS}^2}\right]^{1/4}.
    \end{equation}
In our simulated spectrum, the waste heat of the {hot} Dyson Sphere is {$\sim9.5\times10^{4}\,{\rm K}$.}
We assume that with the range of $\eta=[0.01,0.99]$, Eddington ratio $=[0.1\%,100\%]$ and radius of the Dyson Sphere $=[10^3R_{\rm Sch},10^5R_{\rm Sch}]$.
The possible wavelength (peak wavelength in the blackbody spectrum of the waste heat) to detect {a hot} Dyson Sphere as a function of the black hole mass is shown in Fig. \ref{figwav}.
Our result suggests that the waste heat of a {hot} Dyson Sphere around a black hole is detectable in the UV, optical, near-infrared (NIR), and mid-infrared (MIR) band.
We caution readers that star forming activity and dusty torus would be detected in the longer wavelengths such as IR, which do not disturb the detection of the {hot} Dyson Sphere in the spectrum.
{However, we caution readers that in our simulation, we assume there is no dust extinction.
The waste heat may be severely affected by dust extinction within the Milky Way.
Only if the black hole is close to the Earth or is observed in the infrared wavelength, this problem can be avoided.
Similarly, a Dyson Sphere of more massive black holes are redder, and less affected by the dust extinction. For example, at $M\gtrsim10^{5}-10^{6}\,{\rm M_{\odot}}$, most emissions from Dyson spheres are in the infrared.
}

{In terms of searching for Dyson Spheres around black holes, since Dyson Spheres may emit similar temperatures as stars, UV, optical, and infrared sky surveys might be inefficient.
We recommend analysing black holes (candidates) in the current catalogue first.
If there is spectroscopic data, we can analyse if there are several abnormal peaks.
If only photometric data are available, at least we need several band detection such as UV, optical, and infrared to perform spectral energy distribution (SED) fitting \citep[e.g.,][]{Hsiao2020} with the Dyson Sphere model spectrum, which will help us to identify the possible Dyson Sphere.}

{There are many factors that may affect the detectability.
First of all, if the covering fraction or the transfer efficiency of the Dyson Sphere is too small, the feature might be hidden beneath the disk/companion star spectrum.
As a result, we may mis-identify them as normal X-ray binaries.
In addition, even if the Dyson Sphere is bright enough, it may be challenging to distinguish it from a companion star or an exoplanet orbiting around the X-ray binaries \citep[e.g.,][]{Imara2018}.
We propose to overcome this problem through model fitting (SED fitting). If the blackbody radiation has an abnormal luminosity with a temperature that cannot be explained by usual models in SED fitting (e.g., star formation histories, etc....), we could further consider the Dyson Sphere scenario (e.g., $T=2000\,{\rm K}$).
{More robust confirmation comes with a spectrum by checking the absorption lines since low-temperature stars have spectra that have strong molecular bands.}
If the waste heat of a Dyson Sphere has a temperature below $\sim2000\,{\rm K}$, there would be a clear bump in the infrared spectrum that is not expected from a companion star with molecular bands.
In addition, if a Dyson Sphere collects a huge amount of the energy from the disk, it is possible that the bolometric luminosity is much lower than expected. For example, the spectrum is thermal blackbody radiation but the Eddington ratio is much smaller than $1\%$, which should be a thick disk and shows a non-thermal feature in the spectra \citep[e.g.,][]{Shakura1973,Narayan1994}.
This may indicate that the real Eddington ratio is higher but a fraction of the energy is absorbed by a Dyson Sphere.
Moreover, a Dyson Sphere should have a much smaller mass than a companion star or an exoplanet.
Operating a telescope to measure the variability of radial velocity will also help distinguishing between companion stars/exoplanets and Dyson Spheres candidates.
}


\citet{Inoue2011} discussed the energy transmitted remotely from a Dyson Sphere to a Type II civilisation's habitat.
In their result, the possibility to detect the transmission of a $1\,{\rm \mu arcsec}$ energy beam is less than $10^{-23}$.
For a stellar-mass black hole, the possibility becomes smaller.
A search for these signals from a black hole in the UV, optical, and IR bands is needed to discover the existence of a Type II civilisation.

    \begin{table*}
    	\centering
    	\caption{Information of the projects.}
    	\label{tab2}
    	\begin{flushleft}
    	\begin{tabular}{|c|c|c|c|c|}\hline
project                                                                  & band & exposure time(sec)    & limiting magnitude($m_{\rm AB}$) & reference \\ \hline
WFC3/UVIS of {\it HST} & NUV(F225W) & 3600(1\,hr) & $\sim27$ & $^{a}$ \\ \hline
\multirow{2}{*}{{\it GALEX} Ultra-violet Sky Surveys} & NUV  & \multirow{2}{*}{100}  & 19.9 &\multirow{2}{*}{\citet{Bianchi2011}} \\ 
                                                                             & FUV  &                       & 20.8   \\ \hline
\multirow{5}{*}{Pan-STARRS1 $3\pi$ Steradian Survey}                            & $g_{\rm P1}$  & 43                    & 23.3 & \multirow{5}{*}{\citet{Chambers2016}} \\ 
                                                                             & $r_{\rm P1}$  & 40                    & 23.2  \\ 
                                                                             & $i_{\rm P1}$  & 35                    & 23.1  \\ 
                                                                             & $z_{\rm P1}$  & 30                    & 22.3  \\ 
                                                                             & $y_{\rm P1}$  & 30                    & 21.4  \\ \hline
\multirow{4}{*}{{\it VHS}}                               & $Y$    & \multirow{4}{*}{60}   & 21.2 &\multirow{4}{*}{$^{b}$} \\ 
                                                                             & $J$    &                       & 21.1  \\ 
                                                                             & $H$    &                       & 20.6  \\ 
                                                                             & $K_{s}$   &                       & 20.0  \\ \hline
\multirow{4}{*}{{\it WISE}}                  & $3.3{\rm\,\mu m}$  & \multirow{4}{*}{8.8}  & 18.7 & \multirow{4}{*}{\citet{Mainzer2005}}     \\
                                                                              & $4.7{\rm\,\mu m}$  &                      &18.4       \\  
                                                                             & $12{\rm\,\mu m}$   &                       &16.9       \\  
                                                                             & $23{\rm\,\mu m}$   &                       &15.4       \\ \hline
\multirow{5}{*}{{\it SDSS}}                             & u    & \multirow{5}{*}{53.9} & 22.0 &\multirow{5}{*}{\citet{Abazajian2004}$^{c}$}  \\ 
                                                                             & $g$    &                       & 22.2  \\ 
                                                                             & $r$    &                       & 22.2  \\ 
                                                                             & $i$    &                       & 21.3  \\  
                                                                             & $z$    &                       & 20.5  \\ \hline

        \end{tabular}\\
$^{a}$\url{https:// HST-docs.stsci.edu/wfc3ihb/chapter-6-uvis-imaging-with-wfc3/6-8-uvis-sensitivity}
$^{b}$\url{https://people.ast.cam.ac.uk/~rgm/VHS/v4/PS-VHS-v4p2.pdf}\\
$^{c}$\url{https://www.SDSS.org/dr16/scope/}        
        \end{flushleft}
    \end{table*}

\section{Conclusion}
\label{Conclusion}

In this study, we consider an energy source of a well-developed Type II or a Type III civilisation.
They need a more powerful energy source than their own Sun.
We discuss and conclude that the collectable energy from the CMB at present by the Inverse Dyson Sphere would be too low ($\sim10^{15}\,{\rm W}$).
Next, the Hawking radiation as a source seems to be rather infeasible since the Hawking luminosity cannot provide adequate energy (e.g., for $5\,{\rm M_{\odot}},\:L_{\rm Hawking}\sim10^{-30}\,{\rm W}$ $<<10^{26}\,{\rm W}$ (Type II)).

On the other hand, an accretion disk, a corona, and relativistic jets could be potential power stations for a Type II civilisation.
Our results suggest that for a stellar-mass black hole, even at a low Eddington ratio, the accretion disk could provide hundreds of times more luminosity than a main sequence star.
If a Type II civilisation collects the energy from the accretion disk of a SMBH, the energy could boost the Kardashev index, $K\sim2.9$.
Moreover, the energy reserved in a corona and jets can provide additional energy ($\sim30\%-\sim50\%$ for the corona luminosity and $\sim60\%-\sim80\%$ for the jets luminosity) aside from the accretion disk.
Our results suggest that if a Type II civilisation collects the energy from jets and electromagnetic radiation simultaneously, for a SMBH with a mass similar to Sgr A*, the Kardashev index can reach $\sim 3$.
Overall, a black hole can be a promising source and is more efficient than harvesting from a main sequence star.

We also discuss a possible location of a Dyson Sphere around a black hole.
To absorb the accretion disk luminosity, a Dyson ring or a Dyson Swarm could be a possible structure.
A Dyson Sphere should be located outside of the accretion disk, $\sim10^3R_{\rm sch}-10^5R_{\rm sch}$.
{However, in this region, the hot temperature would melt the solid structure.
In order to avoid melting of the solid Dyson Sphere, the solid Dyson Sphere ($T=3000\,{\rm K}$) should be located at $R_{\rm DS}\gtrsim10^{7}R_{\rm Sch}$ and
$R_{\rm DS}\gtrsim10^{3}R_{\rm Sch}$ for a stellar mass black hole and a SMBH, respectively.}
The size of an accretion disk for a stellar-mass black hole is smaller than the size of the Sun, which means that a Dyson Sphere around the stellar-mass black hole can be smaller than that around the Sun.
In terms of relativistic jets, a possible form of a Dyson Sphere is a Dyson Bubble.
Balancing the pressure and the gravity from the black hole, a light sail could be stationary in space and can continuously collect the energy from the jets.
Light sails also absorb the luminosity from the accretion disk at the same time.
Our results suggest that the best way to place a light sail is close to the origin of the jets, which could save the materials to build a light sail.

Moreover, a {hot} Dyson Sphere around a stellar-mass black hole in the Milky Way ($10\,\rm kpc$ away from us) is detectable in the UV$(\rm 10-400\,{\rm nm)}$, optical$(\rm 400-760\,{\rm nm)}$, NIR($\rm 760\,{\rm nm}-5\,{\rm \mu m}$), and MIR($\rm 5-40\,{\rm \mu m}$), which can be detected by our current telescopes (e.g., WFC3/{\it HST} {and {\it GALEX} survey}).
{For a solid Dyson Sphere, the limiting magnitude of the sky surveys such as Pan STARRS1, VISTA Hemisphere Survey ({\it VHS}) Wide-Field Infrared Survey Explore ({\it WISE}) and Sloan Digital Sky Survey ({\it SDSS}) are smaller than the flux density from the solid Dyson Sphere, which indicates that the solid Dyson Sphere is bright enough to be detected.
In addition, the presence of Dyson Spheres may be imprinted in spectra.
Performing model fitting and measuring the radial velocity will help us to identify these possible artificial structures.}

\section*{Acknowledgements}
{We thank the anonymous referee for useful comments and constructive remarks on the manuscript.
We appreciate Prof. Hsiang-Yi Karen Yang for helpful discussions.}
TG acknowledges the support by the Ministry of Science and Technology of Taiwan through grant 108-2628-M-007-004-MY3.
TH and AYLO are supported by the Centre for Informatics and Computation in Astronomy (CICA) at National Tsing Hua University (NTHU) through a grant from the Ministry of Education of the Republic of China (Taiwan).
AYLO's visit to NTHU was hosted by Prof. Albert Kong and supported through the Ministry of Science and Technology of the ROC (Taiwan) grant 105-2119-M-007-028-MY3.

\section*{Data Availability}
\label{Data availability}
The data underlying this article will be shared on reasonable request to the corresponding author.

\bibliographystyle{mnras}
\bibliography{references}
\bsp	
\label{lastpage}
\end{document}